# Infrastructure Strategy to Enable Optical Communications for Next-Generation Heliophysics Missions

Marta Shelton, Hongbo Li, Daniel Motto, Antti Pulkkinen, Errol Summerlin, Doug Rabin, (NASA Goddard Space Flight Center)

Ryan Rogalin, Abraham Douglas, Stephen Lichten (NASA Jet Propulsion Laboratory)

Mark Storm, Brian Mathason (Fibertek)

Amir Caspi (Southwest Research Institute)

1. **Introduction**

The topics of the Heliophysics Decadal Survey historically include exploration of the characteristics and physics of the interplanetary medium from the surface of the Sun to *interstellar space beyond the boundary* of the heliosphere. It also supports applications in "emerging frontiers" to robotic and human exploration in and *beyond low-Earth orbit.* Emphasis is placed on improving the *space weather* pipeline from basic research to applications to operations to strengthen *forecasting and predictive* capabilities.

To expand frontiers and achieve measurable progress, instruments such as hyperspectral imagers are increased in resolution, field of view, and spectral resolution and range, leading to dramatically higher data volumes. Increasingly, data need to be returned from greater distances, ranging from the Sun-earth L1/L2 points at 1.5 million km, to L4/L5 halo orbits at 1 AU, to several AU in the case of planetary probes. There is also a drive to go beyond single-point measurements to characterizing fields, involving data collected from a constellation or swarm of satellites (e.g., CATSCaNS, MagCon). An elegant solution for returning science data from as many as 36 satellites calls for a novel, hybrid communications solution. Another revolution is through Distributed Spacecraft Missions (e.g., CMO), where multiple satellites work in concert to enable a common science goal, requiring real-time or near real-time sharing of information, best served by optical comm inter-satellite links. **For some of these decadal concepts, optical communication is not "nice to have" but a necessity**. Otherwise, planning decadal investigations which take place in a deep space region will put a stress on the largest (34-meter and 18-meter) RF ground stations. Communicating seven hours each day with RF from a deep space takes away from the science mode time or having seven passes per day from a lower orbit will be operationally stressful. The data rates available for deep-space RF services requires a compromise on the decadal scientific objectives. **Optical communications can significantly reduce resource competition, requiring significantly fewer passes per day and/or shorter overall passes, and thereby enable far greater, transformative science return from individual missions and the capacity to support multiple such missions within a smaller ground network.**

For end-to-end support of decadal research strategies, a communications infrastructure is needed to improve the data generation/data returned ratio and provide mission planners with this information to incorporate into their concept designs. Optical communications must be a part of a comprehensive solution. It can do the heavy lifting for returning large data volumes to Earth at 20x the data rate of traditional RF frequencies (up to Ka band). As evidenced through work in the GSFC Mission Design Lab, due to RF communications constraints from deep space, missions are often limited to returning only a fraction of the data acquired (sometimes as low as 10%). Optical communications also provides superior performance and increased ranges for Inter-satellite Links (ISL) from 2,000 to 10,000 km for Swarms and DSMs. Lastly, the only way to guarantee timely space weather warnings (with a target of 15 minutes latency) is through space relays in MEO or GEO orbits, a strategy which also includes optical communications.

## 2. Technology gap: ground infrastructure

There is a need for *multi-mission* optical ground stations built on standardized equipment to serve the decadal missions and more, to provide a greater return on investment than the existing *single-mission* optical ground stations that have been built and validated by NASA and its partners. Though a commercialization effort is underway for ground stations servicing near Earth orbits, commercial partners do not have interests or investments to service deep space. This is true for RF ground stations (JPL is the only one with 34m and 70m ground assets), as well as optical ground stations. Commercial sub-1-meter optical ground stations serving up to Lunar ranges are coming online in 2022-2024. However, link budget calculations reveal that to reach 267 Mbps, a minimum 2-meter ground telescope is needed. The same 2-meter telescope can also serve a Distant Retrograde Orbit (DRO) at 0.1 AU at 10 Mbps.

**One-size does not fit all when it comes to optical ground aperture; one-modem does not fit all**. As outlined above, a 2m telescope can serve the Sun-Earth Lagrange points L1 and L2 (see Optical Communications Sustainment Study). L4 and L5 (at 100x the range) are better served by a 4-meter telescope (SETH Heliophysics runner-up proposal). JPL is working toward developing an 8-meter mirror segment for use in the center of DSS-23, a 34-meter RF dish to enable communications from ranges such as Mars. When complete, the ground asset should be able to receive RF and optical signals at the same time. The build of this hybrid has been delayed from the original 2024 to now 2032-2035. When completed, NASA will have quantity one large optical ground station (OGS). Weather analysis reveals that one OGS in the CA area would have an average of 85% fair-day availability. With a second OGS in the southern hemisphere (e.g. Australia), the availability goes up to 97%, a great operational improvement.

In the meantime, a 1.3-meter equivalent optical aperture, a predecessor prototype of the planned 8-meter has been open-air tested by JPL on the experimental DSS-13 station and is awaiting **missions of opportunity** to be fully validated. While the commercial landscape for OGS and optical receivers is a bit of a Wild West when it comes to modulation types for fast, gigabit speed modems to receive from LEO orbits, deep space optical modems for the photon starved environment require a Pulse Position Modulation (PPM) scheme supporting different PPM orders, and only NASA/ JPL has operational products for closing deep space optical links. Such exclusion also means that NASA is in a good position to standardize on the JPL designed optical communications ground receiver suite (superconducting nanowire single photon detector, cryocooler, time-to-digital converter etc.). In part for supporting the DSOC demonstration on Psyche, variations of the same design, as of 2022, have been integrated in four places: at the

Palomar observatory, OCTL, a White Sands telescope to serve as OGS for O2O, and a limited version behind the 1.3m experimental hybrid on DSS-13.

## 3. Technical, risk, and cost assessments

Optical communications can be viewed as *ambitious, but certainly realistic,* as evidenced through multiple optical comm demonstrations (LLCD, LCRD, TBIRD, and anticipating O2O, ILLUMA-T and DSOC). Coarse pointing requirements can be met with COTS ACS components, while devices such as fast steering mirrors keep the space-laser on-target with fine precision. Even in the case of an L1 destination, coveted by many Heliophysics observations, halo orbits can be customized such that a Sun keep-out angle (~20 degrees) is observed at all times so as not to damage the ground optics. As far as mitigation of atmospheric turbulence on the ground, companies such as Cailabs are designing turbulence mitigation components. Tilba-ATMO is based on Multi Plane Light Conversion "(MPLC) technology, with applications in fibered telecommunications, high-power laser processing, and most recently free-space optical communication, enabling scalable, low footprint and robust single mode coupling in all optical links for ground stations" (Cailabs 2022). This technology is rolling out first for coherent receivers.

The risk of weather effects occasionally precluding laser acquisition (clouds, fog) can be mitigated two ways: spatial (geographical) diversity with more than one OGS, and operational temporal margin. While with RF, daily downlinks are required for high data volume instruments, the same volume may be downlinked in just one day per week thanks to the much higher data rate made possible by optical comm, allowing scheduling around weather phenomena, or making it easier to double up data volume transmitted during the next available link.

The implementation of Delay/ Disruption Tolerant Networking (DTN) protocol in the communications network layer where optical communication is deployed also provides operational benefits, particularly in application of intersatellite links and/or aggregate data trunk lines and relays. In cases when the complete end-to-end path is available only part of the time, a DTN architecture may deliver data to the end destination sooner than presently used communication schemes, since DTN can move the data along the path hop by hop without waiting for all nodes along the path to be available simultaneously. The implementation of DTN would help a mission maximize its communications resources.

## 4. The international landscape, inter-agency collaborations, public-private relationships, and innovative partnerships

NASA Heliophysics (as well as Astrophysics and Planetary Science) decadal-class investigations cannot ride the coattails of commercial ground investments, as these target Near Earth opportunities and not the 0.1 AU – 1 AU regime. However, OGS does not need to be cost prohibitive. GSFC's Low-Cost Optical Terminal serves as a blueprint for erecting OGS with modified COTS parts (e.g. telescope). "LCOT has a 70-centimeter receive optical telescope, so it's a little smaller than the ground telescopes for LCRD, but comparable to the O2O ground telescope," said H. Safavi. "We're working with them to test our adaptive optics and tracking systems. Our major goal is to test at lunar distances with O2O." (https://esc.gsfc.nasa.gov/news/Low-Cost_Optical_Terminal_Project/)

While a 70 cm telescope is approximately $500 K (PlaneWave), it can be scaled up to a 2m telescope for $2 M. Costs for a fully operational OGS (for deep space) loaded with a PPM receiver is estimated at $17-20 M depending on location, existing infrastructure, and other advancements such as adaptive optics or Wave Division Multiplexing (WDM) to demultiplex multiple bound optical channels for even greater throughput. In this case, a copy of the JPL optical receiver model (O2O, DSS-13, DSOC) is assumed. **We note that $20M is only about 3% of a typical Discovery-class mission and ~1% of a flagship-level mission, adding marginal cost to even a single large mission. Moreover, a one-time investment per OGS can service many future missions.** Thus, the return-on-investment potential is enormous.

A survey of international optical communications efforts reveals that a European Optical Nucleus Network was formed between ESA ESOC, DLR GSOC and KSAT. Parties agree to have an interoperable multi-mission approach based on CCSDS (The Consultative Committee for Space Data Systems) standards. In Australia, the Australian Optical Ground Station Network (AOGSN) is planned to be made up of four ground stations in Western Australia, South Australia, the Australian National University (Australian Capital Territory), and New Zealand. The plan is to tie these stations together to produce a communication network that can support optical, RF, and future quantum communications. Sascha Schediwy, head of the research group responsible for designing and building the WA Optical Ground Station, believes lasers will play a crucial role in the next human missions to the Moon. "It's likely to be how we'll see high-definition footage of the first woman to walk on the Moon," Dr. Schediwy said (abc.net.au). In Japan, the National Institute of Information and Communications Technology (NICT) optical ground station in Japan also received transmission from the SOLISS system by Sony CSL installed on the Kibo's exposed facility on the ISS.

All of the above current (or planned) optical ground assets are 50 cm-80 cm, too small for post-lunar optical communications, thus NASA is uniquely situated to provide the infrastructure investment that enables state-of-the-art decadal science and science data return from deep space, and as such become the leader and following CCSDS recommendations, standardize on optical ground equipment for multi-mission use and long-term compatibility.

Paired with the appropriately sized optical ground infrastructure, commercial partners such as Fibertek (Herndon, VA) are well positioned to provide low SWaP modified-COTS (optimized for deep space) optical space terminals and Pulse Position Modulation (PPM) waveform software-defined optical modems, capitalizing on NASA SBIRs to develop such technology (doi: 10.1117/12.2590244).

5. Infrastructure: an issue of broad concern to the community

There are specific decadal concepts and demonstration proposals, within and beyond the Heliophysics community, for which optical comm will be a necessity or a significant operational improvement. Some of these are shown in Table 1:

Table 1. High data volume or low latency decadal and other proposals

| Concept Name | Discipline | Orbit Range, max | Data Volume to Earth |
|---|---|---|---|
| CMO | Heliophysics | Sun-Earth L1, 1.5 mil km | 23 Tb per day (total) from 2 spacecraft |
| MagCon | Heliophysics | 8, 11, 15 $R_E$ | Aggregate from 36 satellites |
| OST | Astrophysics | Sun-Earth L2, 1.5 mil km | 30 Tb per day |
| COMPLETE | Heliophysics | Sun-Earth L4/L5, 150 mil km + Sun-Earth L1, 1.5 mil km | 233 Gb per day from each of L4 and L1 (only 10% of total data generation) |
| MOST | Heliophysics | Sun-Earth L4/L5, 150 mil km | Multiple satellites and instruments |
| CATSCaNS | Heliophysics | Heliocentric, 0.1 AU | Multiple satellites |
| GEP / PRIMA | Astrophysics | Sun-Earth L2, 1.5 mil km | Compared to RF, increase resolution and data volume by 7x, or reduce comm time 7x |

Optical Communications will enable not just Heliophysics missions with large data rates and volumes, but also human Mars exploration with long link distances, and robotic missions beyond Mars distance with long ranges and difficult mass & power constraints.

## 6. Next Steps

Specific to the issue of ground infrastructure, maintaining NASA interest and strategic investment in the development of 34-meter hybrid RF/optical terminals for DSN is a critical need for decadal mission proposals. Proposal and therefore mission risks associated with the availability of communication services often weigh heavily on mission selection criteria and are sought to be retired or mitigated during pre-formulation. The current commitment to a progressive development plan of demonstration assets (DSS-13), culminating with an operational capability of northern (DSS-23) and southern (DSS-33) hemisphere optical terminals in the 2030s timeframe, provides some level of mitigation necessary for Principal Investigators and mission planners to have confidence during proposal development. However, to facilitate missions requiring optical communications in the near term, we strongly recommend that NASA expedite deployment of the 8-meter hybrid solution by 2030. Multiple Decadal mission concepts for solar observation from deep space have proposed launches around 2032, and this would provide sufficient time to deploy and test this solution prior to their launch. As a stopgap, a 4-meter version of this asset could be deployed on a shorter timeframe, eliminating the need to repurpose other assets such as Palomar where competition for scientific astronomy would significantly complicate operations.

For these reasons, maintaining or accelerating the commitment to the DSN hybrid antenna project together with adding other medium sized (2-4 meter) optical ground telescopes to form an Optical Ground Station Network is our highest recommendation. A secondary recommendation is for NASA to investigate options for the establishment of the planned 1.3m and/or 4.1m DSS-13 demonstration assets to be used for long-term operational mission use.

## Acronyms

| | |
|---|---|
| ACS | Attitude Control System |
| AU | Astronomical Unit |
| CCSDS | The Consultative Committee for Space Data Systems |
| COTS | Commercial Off-The-Shelf |
| DRO | Distant Retrograde Orbit |
| DSM | Distributed Spacecraft Mission |
| DSN | Deep Space Network |
| DSOC | Deep Space Optical Communications |
| DSS | Deep Space Station |
| DTN | Disruption Tolerant Networking |
| RF | Radio Frequency |
| ILLUMA-T | Integrated LCRD LEO User Modem and Amplifier Terminal |
| ISL | Inter-satellite Links |
| LEO | Low Earth Orbit |
| LCOT | Low Cost Optical Terminal |
| LCRD | Laser Communications Relay Demonstration |
| LLCD | Lunar Laser Communication Demonstration |
| MEO | Medium Earth Orbit |
| GEO | Geostationary Earth Orbit |
| O2O | The Orion Artemis II Optical Communications System |
| OCTL | Optical Communications Telescope Laboratory |
| OGS | Optical Ground Station |
| PPM | Pulse Position Modulation |
| SETH | Science-Enabling Technologies for Heliophysics |
| TBIRD | TeraByte InfraRed Delivery |
| Tx | Transmitter |
| WDM | Wave Division Multiplexing |